\documentclass[reprint,superscriptaddress,aps,prd,floatfix]{revtex4-2}
\usepackage{placeins}
\usepackage[a4paper,left=1.5cm,right=1.5cm,top=3cm,bottom=3cm]{geometry}
\usepackage[toc]{appendix} 

\usepackage{graphicx}
\usepackage{dcolumn}
\usepackage{bm}
\usepackage{xcolor}
\usepackage{amssymb,amsmath,amsfonts,physics}
\usepackage[dvipsnames]{xcolor}
\usepackage{graphicx}
\usepackage{longtable}
\usepackage{verbatim}
\usepackage{color}
\usepackage{mdframed}
\usepackage{soul}
\usepackage{amsfonts,amssymb,mathrsfs,amsmath,esint}
\usepackage{slashed, cancel}
\usepackage{framed}
\usepackage{mdframed}
\usepackage{simplewick} 
\allowdisplaybreaks
\usepackage{latexsym}
\usepackage{graphicx}
\graphicspath{{./Figures/}}
\usepackage[dvipsnames]{xcolor}
\usepackage{booktabs}
\usepackage{datetime}
\usepackage{float}
\newdateformat{mydate}{\THEDAY{ }\monthname[\THEMONTH]{ }\THEYEAR}

\usepackage{tikz}
\usepackage{color}
\usepackage{framed}
\usepackage{hyperref}
\usepackage[capitalise]{cleveref}
\definecolor{rossos}{cmyk}{0,1,1,0.55}
\definecolor{bluscuro}{rgb}{0.15, 0.2, .85}
\definecolor{bluchiaro}{cmyk}{1,.3,0.,0.1}
\hypersetup{colorlinks, citecolor=bluscuro, linkcolor=black, urlcolor=bluscuro}
\definecolor{rossos}{cmyk}{0,1,1,0.55}
\definecolor{bluscuro}{rgb}{0.15, 0.2, .85}
\definecolor{bluchiaro}{cmyk}{1,.3,0.,0.1}

\graphicspath{{./images/}}

\newcommand{\be}{\begin{equation}}
	\newcommand{\ee}{\end{equation}}
\newcommand{\bea}{\begin{eqnarray}}
	\newcommand{\eea}{\end{eqnarray}}
\newcommand{\beq}{\begin{equation}}
	\newcommand{\eeq}{\end{equation}}

\def\beqa{\begin{eqnarray}}

	\def\eeqa{\end{eqnarray}}

\def\lsim{\mathrel{\rlap{\lower4pt\hbox{\hskip0.5pt$\sim$}}
		\raise1pt\hbox{$<$}}}         
\def\gsim{\mathrel{\rlap{\lower4pt\hbox{\hskip0.5pt$\sim$}}
		\raise1pt\hbox{$>$}}}         

\usepackage[normalem]{ulem}
\usepackage{soul}

\def\be{\begin{equation}}
\def\ee{\end{equation}}
\def\bea{\begin{eqnarray}}
\def\eea{\end{eqnarray}}
\def\beaN{\begin{eqnarray*}}
\def\eeaN{\end{eqnarray*}}
\def\ed{\end{document}}
\def\bit{\begin{itemize}}
\def\eit{\end{itemize}}

\def\~{\tilde}

\newcommand{\ns}{n_{s}}
\newcommand{\rr}{r}
\newcommand{\Nstar}{N_{\star}}
\newcommand{\lnB}{\ln B_{01}}
\newcommand{\Mzero}{\mathcal{M}_{0}}
\newcommand{\Mone}{\mathcal{M}_{1}}

\usepackage[dvipsnames]{xcolor}
\usepackage{graphicx}

\def\be{\begin{equation}}
\def\ee{\end{equation}}
\def\bea{\begin{eqnarray}}
\def\eea{\end{eqnarray}}

\begin{document}

\nocite{OoguriVafaSwampland,Garg:2018reu,Agrawal:2018own,BenDayan2019DrainingSwampland,Artymowski:2019vfy,Bezrukov:2014bra,Hamada:2014iga,Rubio:2015zia,Bezrukov:2014ipa,Masina:2018ejw,Salvio:2017oyf,Masina:2024ybn,Ben-Dayan:2008fhy,Ben-Dayan:2009elk,Wolfson:2022ckx}

\title{Slowly Rolling on a Quantum Correction}

\author{Ido Ben-Dayan}
\affiliation{Physics Department, Ariel University, Ariel 40700, Israel}

\author{Ayushi Srivastava}
\affiliation{Physics Department, Ariel University, Ariel 40700, Israel}

\author{Amresh Verma}
\affiliation{Physics Department, Ariel University, Ariel 40700, Israel}

\begin{abstract}
{Recent advances in cosmological measurements such as DESI and ACT may indicate an increase in the preferred value of the spectral tilt $n_s$ such that it disfavors rather popular models such as the Higgs or Starobinski inflation models. We argue that it actually means that the data is now sensitive enough to quantum corrections beyond simple tree-level models. Resurrecting the old theme of Coleman-Weinberg effective potential, we analyze the predictions of such models, as well as likelihood analysis, leading to a more established and interesting predictive framework.}
\end{abstract}

\maketitle

\section{Introduction}
Inflation is currently the leading candidate of explaining the observed temperature anisotropies of the cosmic microwave background radiation (CMB) as well as the flatness and horizon problems in the Hot Big Bang Cosmology \cite{Planck2018Inflation}. A full fledged model should address the initial conditions at the onset of inflation, its evolution, and the process of populating the universe with Standard Model (SM) particles, usually dubbed ``reheating" \cite{Ben-Dayan:2023lwd,Ben-Dayan:2025bqd}. However, part of the success of Inflation is that one reaches the inflationary phase from rather generic initial conditions, and that given such a phase of Inflation, one can induce reheating rather generically. This allows model builders to compartmentalize model building, and focus on the evolution of the Universe during Inflation. While Inflation can be driven by multiple various degrees of freedom, the essentials can be reduced to a single canonically normalized, minimally coupled scalar field slowly-rolling down a potential $V(\phi)$. After imposing the slow-roll conditions $\varepsilon,\eta \ll 1$, the complicated multi-faceted theory of Inflation becomes a simple problem of calculus, where the main challenge is to find justification for considering some specific potential. In principle, the potential $V(\phi)$ should be the full quantum-corrected one, after taking into account all possible quantum corrections including ones coming from quantum gravity (QG). Barring a fully constructed model, say from string theory which we currently do not possess, QG can destroy the slow-roll behavior since we expect $\Delta \eta_{QG} \sim 1$. However, we can take into account quantum corrections coming from field theory calculations $\delta V\sim {V_0^{\prime\prime}}^2 \ln V_0^{\prime\prime}$, where $V_0$ is the tree level potential of the model\footnote{Another virtue of logarithmic potentials is that they come as close as possible to saturate the Swampland Conjecture~\cite{OoguriVafaSwampland,Garg:2018reu,Agrawal:2018own, BenDayan2019DrainingSwampland, Artymowski:2019vfy}).}. Given that in the slow-roll regime $\eta=V^{\prime\prime}/V\,$, these corrections are expected to be small, and given the accuracy of data in the past decades they were conveniently neglected, paying lip service to the idea that the ascribed potential is actually a ``quantum corrected" one. Given the increasing accuracy of data, and specifically the combination of ACT and DESI data show that this approach is no longer tenable \cite{DESI:2025zgx}. The spectral tilt is specified to the level of $\mathcal{O}(10^{-4})$, and the error bars are of the order of $\mathcal{O}(10^{-3})$. With such accuracy ${V_0^{\prime\prime}}^2\ln V_0^{\prime \prime}$ may be relevant. One way to address the new results is to construct new ``tree-level" potentials that better fit the data e.g. \cite{Ben-Dayan:2013fva, Artymowski:2019jlh}. However, unless calculated, these new models could again have quantum corrections that may quantum correct them away from the desired values. Hence, a better way to proceed is to consider existing models, explicitly evaluate their quantum corrected potential, and only then apply the slow-roll conditions and analyze their predictions. We demonstrate this idea in three explicit examples, the Starobinski model, a self-interacting scalar field non-minimally coupled to gravity, and the SM Higgs Inflation model. In all examples, the result is a sizeable enough observable modification of the spectral tilt $n_s$ and tensor-to-scalar ratio $r$. In the first two examples the quantum corrected versions interpolate between something like their tree-level version $V\sim (1-e^{-a\chi})^2$, and a linear potential $V\sim \chi$. For the SM Higgs, due to the beta functions the correction goes the other way, the point being, \textit{the data is accurate enough. For accurate predictions, one must really calculate the quantum corrected potential and derive predictions from it.}
We present the models in their quantum corrected version, and calculate for each the resulting shifted observables. We then perform a likelihood analysis, and demonstrate how the quantum corrected version is a better or worse fit to the data.
In all examples, the action is first specified in the Jordan frame, transformed to the Einstein frame, and then we canonically normalize the scalar field.
Throughout the manuscript we will be using natural units, $M_{pl}^2\equiv (8\pi G)^{-1}\equiv 1$, the standard observables $n_s = 1+ 2\eta- 6\epsilon,\,
r = 16\epsilon$,  and slow-roll expressions for a canonical scalar field, $\chi$:
\begin{equation}
\epsilon
=
\frac{1}{2}
\left(
\frac{1}{U}\frac{dU}{d\chi}
\right)^2,
\quad 
\eta
=
\frac{1}{U}\frac{d^2 U}{d\chi^2}, \quad
N
=
\int_{\chi_{\text{end}}}^{\chi}
\frac{U}{dU/d\chi}\, d\chi.\nonumber
\end{equation}

\section{Models}
\textbf{1-Loop Corrected Starobinsky Inflation:}
The 1-loop Starobinsky gravity lagrangian density is
\begin{equation}
f(R)=R+\alpha R^{2}+\beta R^{2}\ln R ,
\label{eq:fR}
\end{equation}
where \(\alpha\) is the usual Starobinsky coupling and \(\beta\) is usually assumed to be negligible. The full analysis was carried out in \cite{Ben-Dayan:2014isa}, see also \cite{Codello:2014sua}.
The Einstein-frame scalar degree of freedom is introduced through the conformal transformation
$e^{\tilde{\chi}} \equiv f'(R)
=1+(2\alpha+\beta)R+2\beta R\ln R$ ,
with $\chi=\sqrt{\frac32}\,\tilde{\chi}$, which maps the theory to standard Einstein gravity plus a canonical scalar field.
For phenomenological applications the regime \(|\beta| \ll \alpha\) is sufficient, with the curvature and potential approximated
\bea
R(\chi)&\simeq&
\frac{e^{\tilde{\chi}}-1}
     {2\alpha}
\left[
1+\frac{\beta}{\alpha}
\left(
\frac12+
\ln\!\frac{e^{\tilde{\chi}}-1}{2\alpha}
\right)
\right]^{-1},
\label{eq:Rapprox}\\
U(\tilde{\chi})&\simeq&
\frac{
U_s(\tilde \chi)
}{
1+\dfrac{\beta}{\alpha}
\left[
\frac12+
\ln\!\left(
\frac{e^{\tilde{\chi}}-1}{2\alpha}
\right)
\right]
},
\label{eq:corrected_potential}
\eea 
which is a good approximation as long as 
$\ln(R/\mu^{2}) \sim \ln \frac{e^{\tilde{\chi}}-1}{2\alpha} \sim \mathcal{O}(1)$ across the observable window (the standard Coleman--Weinberg
scale choice). 
Only the ratio
$\frac{\beta}{\alpha}\equiv -t$ affects the observables $(\ns,\rr)$. 
Our implementation reproduces the analytic Starobinsky limit
($\beta \rightarrow 0,\, (\ns)_s=1-2/N$, $\rr_s=12/N^{2}$) to better than $10^{-3}$.
Useful approximate formulae are:
\bea
n_s&=&(\ns)_s+\delta \sqrt{\frac{16 \epsilon_s}{3}},\, \quad
r=r_s+16\left(-\delta\sqrt{\frac{4 \epsilon_s}{3}}+\frac{\delta^2}{3}\right)\cr
N &\simeq & -\frac{3 \log \left[\frac{1-\frac{\delta^2}{3\epsilon_s}}{1-\frac{\delta^2}{3}}\right]-6 \text{Arctanh}\left[\frac{\delta}{\sqrt{3\epsilon_s}}\right]}{\delta(2+\delta)},
\eea
where the subscript $s$ denotes the original tree-level Starobinsky values, and $\delta=\frac{\beta/\alpha}{1+\frac{\beta}{\alpha}\left(1/2+\ln[(e^{\tilde \chi}-1)/2\alpha]\right)}$.\\

\textbf{Self-interacting renormalizable scalar field non-minimally coupled to gravity:}
\label{sec:nonmintheory}
The tree-level Jordan frame action is
\begin{equation}
S_J = \int d^4x \sqrt{-g}\left[
\frac{1}{2}\left(1 + \xi\,\phi^2\right) R
-\frac{1}{2}(\partial\phi)^2
- V_0(\phi).
\right].
\end{equation}
For a given tree-level potential $V_0(\phi)$, we can write the n-loop or RG improved potential that takes into account the running of the couplings. At each given order they should yield the same result.
For example, the 1-loop effective potential for a single scalar degree of freedom is
\begin{equation}
V_{1-loop}(\phi,\mu)
=
V_0(\phi)
+
\frac{1}{64\pi^2} V_0^{"2}(\phi)
\left[
\ln\!\left(\frac{V_0^"(\phi)}{\mu^2}\right) - \frac{3}{2}
\right],
\end{equation}
For simplicity, we consider $V_0\simeq \frac{\lambda}{4}\phi^4$ and neglect other operators. At large field $\phi \gg 1$ this simplifies to 
\be
V_{1-loop}(\phi \gg1) \simeq \frac{\lambda}{4}\phi^4+\frac{9\lambda^2}{32 \pi^2}\phi^4\ln \phi=\frac{\lambda}{4}\phi^4\left(1+\frac{9 \lambda}{8\pi^2} \ln \phi \right)
\ee
Alternatively, promoting couplings to running quantities and choosing $\mu = \phi$, the RG-improved potential is
$V_{\text{RG}}(\phi)
=
V\big(\phi;\lambda(\phi),\xi(\phi)\big).
$ 
Defining the conformal factor and Einstein frame metric
$\Omega^2(\phi) = 1 + \xi(\phi)\phi^2,
\, g_{\mu\nu}^{E} = \Omega^2(\phi)\, g_{\mu\nu}^{J},
$ 
yields the Einstein frame action
\begin{equation}
S_E = \int d^4x \sqrt{-g_E}
\left[
\frac{1}{2}R_E
-\frac{1}{2}K_E(\phi)(\partial\phi)^2
- V_E(\phi)
\right],
\end{equation}
with the kinetic term and potential: 
\begin{equation}
K_E(\phi)
=
\frac{1
+
6 \left(\frac{d\ln\Omega}{d\phi}\right)^2}{\Omega^2}
,\quad 
V_E(\phi)=\frac{V_{\text{RG}}(\phi)}{\Omega^4(\phi)}.
\end{equation}
We then canonically normalize the scalar field $\chi$,
\begin{equation}
\frac{d\chi}{d\phi}
=
\sqrt{
\frac{1}{\Omega^2}
+
\frac{6\xi^2(\phi) \phi^2}{\Omega^4}},
\end{equation}
where we included the possible running of $\xi$.
The inflationary potential in terms of the canonical field is
\begin{equation}
U(\chi)
=
\frac{
V\big(\phi(\chi);\lambda_i(\phi(\chi))\big)
}{
\left(1 + \xi(\phi(\chi))\,\phi^2(\chi)\right)^2
}.
\end{equation}
For large field $\sqrt{\xi} \phi \gg 1$ we can simplify this expression considerably since
$\chi \simeq \sqrt{6} \ln \phi/\phi_0, \quad \phi\simeq \phi_0 e^{\chi/\sqrt{6}} \quad \Rightarrow \Omega^2\simeq \xi \phi_0^2 e^{2\chi/\sqrt{6}}
$. 
Picking $\phi_0=\xi^{-1/2}$, 
the potential at large field is:
\begin{equation}
\begin{aligned}
U_{1\text{-loop}}(\chi)
&\simeq
\frac{\lambda_0}{4\xi_0^2}
(1-e^{-2\chi/\sqrt6})^2
\left[1+\frac{9\lambda_0}{8\pi^2}
\left(\frac{\chi}{\sqrt6}-\frac12\ln\xi_0\right)\right],
\end{aligned}
\end{equation}
where we clearly denote $\lambda_0,\xi_0$ taken at their tree-level value, such that these are truly constants and at this level of accuracy do not depend on $\chi$ anymore. Denoting $c=\frac{9\lambda}{8\sqrt{6} \pi^2}$, the constant term can be absorbed into $\tilde \xi=\xi_0/\sqrt{1-9\lambda/16 \pi^2 \ln \xi_0}$ simplifying the potential,
\be
U_{1-loop}(\chi)\simeq \frac{\lambda_0}{4\tilde \xi^2 }\left(1-e^{-\sqrt{2/3} \chi}\right)^2\left[1+c \,\chi\right].
\ee
We see that the 1-loop correction induces a linear term that multiplies the tree-level potential, as well as renormalizes $\xi\rightarrow \tilde \xi$. Thus, the above potential is the starting point of the slow-roll analysis. 
Evaluating the inflationary predictions directly yields for $c=0.1$, $\chi_{50}\simeq 7.3,\, \chi_{60}\simeq 7.9$ and the corresponding observables
$n_s\in [0.983,0.986],\, r\in [0.031,0.027]$,
respectively,
while for a more reasonable choice of $c=0.01$, $\chi_{50}\simeq 5.5,\, \chi_{60}\simeq 5.77$
\be
n_s\in [0.968,0.975], \quad r\in [6.3,4.7] \times 10^{-3}
\ee
In any case, it is clear that the inclusion of such quantum correction accommodates higher $n_s$, improves the fit to data that includes the DESI results, and can be detected by the upcoming CMB experiments. 

\textbf{The SM Higgs:}
The most constrained model of inflation is arguably the Higgs Inflation model, as any change in the model parameters to fit inflation, can modify collider experiments and vice versa. Specifically, effects due to quantum corrections are highly constrained.  Various analyses including loop corrections, have been studied
extensively~\cite{Bezrukov:2007ep,Bezrukov:2008ej,Bezrukov:2009db, Ben-Dayan:2010vsj,Bezrukov:2010jz,Fumagalli:2016lls,Allison:2013uaa,Rubio:2018ogq,Rodrigues:2021txa}, and we shall follow the expressions of \cite{Allison:2013uaa} throughout. Also relevant is the stability of the SM, assuming no New Physics until the Planck scale. The Higgs+gravity sector action in the Jordan frame,
the transformation to the Einstein frame and canonical normalization are similar the ones described in the self-interacting scalar field section, as well as the tree-level predictions of $n_s\simeq 1-2/N$ and $r\simeq 12/N^2$. The difference with respect to the self-interacting scalar is that the Higgs is part of a doublet and has SM quantum numbers, which dictates the RG flow and the resulting quantum corrections. 
We adopt prescription~II in \cite{Allison:2013uaa},  
where the 1-loop correction is computed in the Jordan frame, so that loop masses scale as $ M^2 \propto h^2$ 
with particle masses,
\begin{align}
\label{e:masses}
M_W^2 &= \frac{g^2 h^2}{4},\,
M_Z^2 = \frac{(g^2+g'^2)h^2}{4}, \,
M_t^2 = \frac{y_t^2 h^2}{2},
\nonumber \\ 
M_h^2 &= 3s\lambda h^2, \, 
M_G^2 = \lambda h^2 .
\end{align}
%
After substitution, the RG improved potential becomes 
\begin{multline}
\label{e:effepot}
U_{RG}(\chi(h)) = \frac{h^4}{4\left(1 + \xi h^2/M_P^2\right)^2}
\Bigg\{
\lambda
+ \frac{1}{16\pi^2}\bigg[
\frac{3g^4}{8}\Big(\ln\frac{g^2}{4} - \frac{5}{6}\Big) \\
+ \frac{3(g^2+g'^2)^2}{16}\Big(\ln\frac{g^2+g'^2}{4} - \frac{5}{6}\Big)
- 3y_t^4\Big(\ln\frac{y_t^2}{2} - \frac{3}{2}\Big) \\
+ 9s^2\lambda^2\Big(\ln(3s\lambda) - \frac{3}{2}\Big)
+ 3\lambda^2\Big(\ln\lambda - \frac{3}{2}\Big)
\bigg]
\Bigg\},
\end{multline}
where $\lambda \equiv \lambda(\mu)$, $\xi \equiv \xi(\mu)$,
$g,g',y_t$ are evaluated at $\mu = h$, and obey the
renormalization group equations \cite{Allison:2013uaa}. 
The quantity $s$ is a suppression factor,
which multiplies each off-shell physical-Higgs propagator, reducing to $s\to 1$ at the electroweak scale and to $s\to 1/(1+6\xi)$ during inflation.

\begin{table}[h]
\centering
\caption{SM couplings at $\mu = M_t$ from
Ref.~\cite{Allison:2013uaa}, for $M_h = 125$~GeV,
$M_t = 170.25$~GeV, $\alpha_s(M_Z) = 0.1184$.}
\label{tab:sm}
\begin{tabular}{ccccc}
\hline\hline
$\lambda$ & $y_t$ & $g$ & $g'$ & $g_s$ \\
\hline
$0.1259$ & $0.9199$ & $0.6483$ & $0.3587$ & $1.1641$ \\
\hline\hline
\end{tabular}
\end{table}
The SM couplings at $\mu=M_t$ are listed in Table~\ref{tab:sm}, which we evolve to all other scales using the RG equations. Unlike the SM couplings, $\xi$ has no measured value and is instead fixed by the CMB normalization 
$\frac{U(h_*)}{\epsilon(h_*)} \simeq 24\pi^2 A_s$,
where $A_s=2.099\times 10^{-9}$ is the measured scalar amplitude \cite{louis25}. Because  $A_s \propto \lambda/\xi^2$,
while both $U$ and $\epsilon$ are evaluated at the horizon-exit field value
$h_*$, which itself depends on $\xi$, 
$\xi$ is determined iteratively, ~\cite{Allison:2013uaa}: For each trial value $\xi(M_t)$ we integrate the
RG equations from $\mu = M_t$, determine the end of inflation from
$\epsilon(h_{\rm end}) = 1$ and the horizon-exit scale $h_*$ from the $e$-fold
number $N$, and adjust $\xi(M_t)$ until the spectral amplitude constraint is satisfied.
For $M_t=170.25$ and $N = 50$--$60$, gives $\xi(M_t) = 2312$--$2847$ with 
\begin{equation}
n_s \in [0.959,\,0.965]\,, \qquad r \in [0.0032,\,0.0021]\,,
\end{equation}
close to the Planck values and below the ACT central value \cite{louis25}.
Following the recent ACT results, several studies have investigated the impact of radiative corrections on inflationary predictions, finding that quantum corrections can shift the scalar spectral index $n_s$ toward the ACT-preferred range \cite{Gialamas:2025kef,Han:2025cwk,Ellis:2025bzi,Ahmed:2025rrg,Wolf:2025ecy,Yuennan:2025kde}. In these treatments strength of the correction is left as a free parameter, whereas in our model we compute the full 1-loop SM potential with the couplings fixed at their measured values. Contrary to the other two examples, the resulting correction in the SM lowers $n_s$ compared to its tree-level value. This suggest that SM Higgs inflation with genuine radiative corrections is
disfavored by current data.\footnote{One possible loophole is 
the so called critical Higgs model \cite{Bezrukov:2014bra,Hamada:2014iga,Rubio:2015zia,Rubio:2018ogq,Bezrukov:2014ipa,Masina:2018ejw,Salvio:2017oyf,Masina:2024ybn}, which is an inflection type model of the form \cite{Ben-Dayan:2008fhy,Ben-Dayan:2009elk,Wolfson:2022ckx}. This freedom allows covering a wide parameter range of observables.}

\section{Results}
\label{sec:results}

We confront each model with the publicly released, fully marginalized $(\ns,\rr)$
posteriors of two dataset combinations~\cite{Balkenhol2025}: \textbf{CMB} (SPA+BK; the SPT+Planck+ACT
`SPA' combination of primary CMB and lensing~\cite{camphuis25,%
louis25,naess25,calabrese25,planck18-1,carron22,qu24,madhavacheril24,ge24,quan26,%
balkenhol24}, plus BICEP/Keck~\cite{BK18}) and
\textbf{CMB+DESI} (adding DESI BAO~\cite{DESI:2025zgx}). Since the models modify only
the primordial sector, the marginalized $(\ns,\rr)$ posterior is a sufficient
statistic, and we build the likelihood as a Gaussian KDE of that posterior,
evaluated on the model prediction interpolated from precomputed grids, discarding other potential observables \cite{Chluba:2012we,Ben-Dayan:2013eza,Ben-Dayan:2014iya,Ben-Dayan:2015zha,Ben-Dayan:2019gll,Ben-Dayan:2023lwd,Ben-Dayan:2024uvx,Ben-Dayan:2025bqd}. With priors specified in Appendix \ref{app:method}, we assess each deviation from
the tree-level model with four complementary criteria, a profile $\Delta\chi^{2}$, the
Savage--Dickey Bayes factor, the Bayesian evidence (dense quadrature, cross-checked
by nested sampling), and the prior-independent deviance information
criterion, with positive $\lnB$ or $\Delta\mathrm{DIC}$ favoring the tree-level model, see Appendix~\ref{app:method} for details.

\begin{figure}[t]
\centering
\includegraphics[width=\columnwidth]{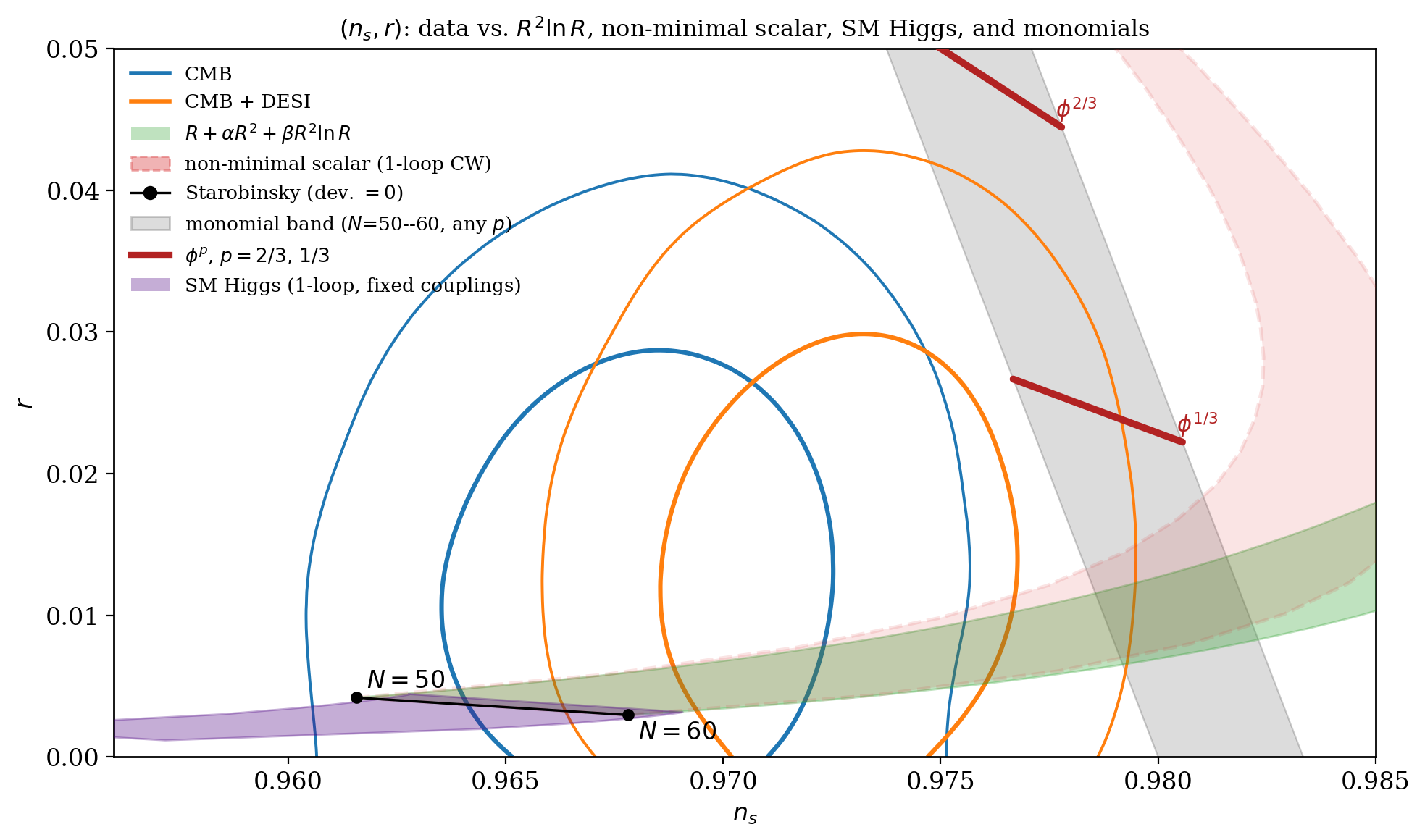}
\caption{Predicted $(\ns,\rr)$ tracks of the models ($50\le\Nstar\le60$) over
the {\it full} deviation range, on CMB (blue) and CMB+DESI (orange) contours:
$R^{2}\ln R$ (green), non-minimal scalar (red) and SM Higgs (indigo). All emanate from the
"Starobinsky segment" (black). 
The grey band is the single-field monomial family
$V\propto\phi^{p}$ swept over $\Nstar=50$--$60$, with the reference potentials
$p=2/3,\,1/3$ marked (dark-red segments).}
\label{fig:widecontour}
\end{figure}

\textbf{Parameter constraints:}
The $R^{2}\ln R$ and  non-minimal scalar constraints are summarized in Table~\ref{tab:modelcomp}, with full posteriors in 
Appendix~\ref{app:post}. With CMB alone the correction strength rails to the
Starobinsky boundary, $t<0.017$ (95\% C.L.), with $\ns=0.9692^{+0.0026}_{-0.0030}$,
$\rr=4.9^{+0.7}_{-1.4}\times10^{-3}$. Adding DESI raises $\ns$ to $0.9729\pm0.0029$
and $\rr$ to $6.0^{+1.2}_{-1.9}\times10^{-3}$ and moves $t$ to
$t=0.0135\pm0.0059$ 
excluding zero at $95\%$ C.L.
The non-minimal scalar gives a parallel constraint:
$c<0.018$ (CMB) and $c=0.0143\pm0.0069$ (CMB+DESI), with
$\ns=0.9695\pm0.0029$, $\rr=(5.1\pm1.4)\times10^{-3}$ and $\ns=0.9733\pm0.0029$,
$\rr=(6.6\pm1.9)\times10^{-3}$ respectively. 
Figure~\ref{fig:widecontour} shows both models' $(\ns,\rr)$ tracks against the data:
the CMB+DESI contour sits in the $t>0$ (equivalently $c>0$) region while CMB alone straddles the
Starobinsky line. 

\begin{table*}[t]
\centering
\caption{The three models versus tree-level Starobinsky
($\Mzero$: deviation $=0$), with comparisons are relative to $\Mzero$:
$\Delta\chi^{2}>0$ favors the models, while $\lnB>0$ and $\Delta\mathrm{DIC}>0$ favor
Starobinsky. For $R^{2}\ln R$ and non-minimal models, 
deviations are 95\% upper limits for
CMB, and posterior means for CMB+DESI, evidences are from dense
2D quadrature, and the $\Delta\chi^{2}$ significance is one-sided (boundary
$\tfrac12\chi^{2}_{0}+\tfrac12\chi^{2}_{1}$). For the SM the Deviation row gives the joint-posterior inferred $M_t$ (in GeV), and there is no nested $\Delta\mathrm{DIC}$. Its
$\Delta\chi^{2}$ and $\lnB$ are evaluated in the joint $(\ns,M_t)$ plane.}
\label{tab:modelcomp}
\small
\setlength{\tabcolsep}{8pt}
\begin{tabular}{lcccccc}
\toprule
 & \multicolumn{2}{c}{$R^{2}\ln R$ ($t$)} & \multicolumn{2}{c}{non-minimal ($c$)} & \multicolumn{2}{c}{SM Higgs ($M_t$)} \\
\cmidrule(lr){2-3}\cmidrule(lr){4-5}\cmidrule(lr){6-7}
\textbf{Quantity} & CMB & +DESI & CMB & +DESI & CMB & +DESI \\
\midrule
Deviation ($t$,\,$c$) / $M_t$~[GeV] & $<0.017$ & $0.0135$ & $<0.018$ & $0.0143$ & $170.0$ & $169.8$ \\
$\Delta\chi^{2}=\chi^{2}_{\Mzero}-\chi^{2}_{\rm model}$ & $0.53$ & $3.48$ & $0.54$ & $3.51$ & $-4.7$ & $-4.9$ \\
significance & $0.7\sigma$ & $1.9\sigma$ & $0.7\sigma$ & $1.9\sigma$ & $2.3\sigma$ & $3.0\sigma$ \\
$\lnB$ & $+0.60$ & $-1.75$ & $+0.44$ & $-2.04$ & $+4.1$ & $+4.3$ \\
$\Delta\mathrm{DIC}$ & $+0.20$ & $-2.84$ & $+0.25$ & $-3.04$ & --- & --- \\
\bottomrule
\end{tabular}
\end{table*}

\textbf{Evidence for a deviation from tree-level:}
The four statistics are collected in Table~\ref{tab:modelcomp}. For CMB alone
every measure is null: $\Delta\chi^{2}=0.53$ ($0.7\sigma$),
$\lnB=+0.60$ (mild, inconclusive preference for Starobinsky/tree-level), and
$\Delta\mathrm{DIC}=+0.20$. For CMB+DESI all four point the same way and at
comparable strength: $\Delta\chi^{2}=3.48$ ($\sim1.9\sigma$, one-sided),
$\lnB=-1.75$ ($\simeq6{:}1$ odds for the extension, though ``weak'' on the Jeffreys
scale), and $\Delta\mathrm{DIC}=-2.84$ (``positive'').
The three $\lnB$ estimators, Savage--Dickey, quadrature, and nested
sampling, agree to $\lesssim0.15$, and the DIC points the same way, so the verdict is method-independent. All prefer the inclusion of quantum corrections.
The DESI hint is thus {\it robust to the deformation mechanism}, while the data
cannot identify which underlying theory is responsible which is an expected consequence of
the tree-level plateau degeneracy. The two models coincide inside the data region
and diverge only at large, non-perturbative deviation (Fig.~\ref{fig:widecontour}).

\begin{figure}[t]
\centering
\includegraphics[width=\columnwidth]{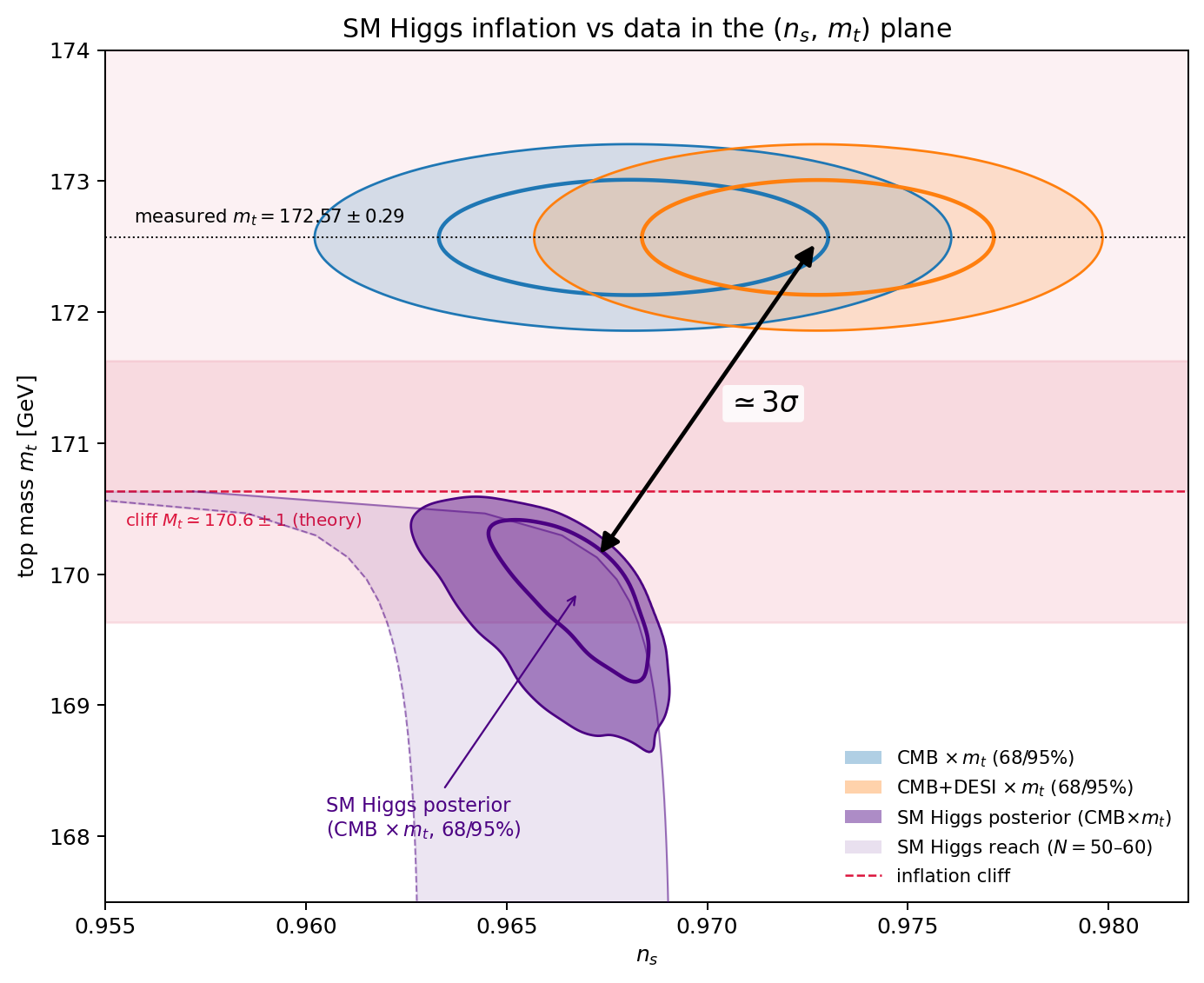}
\caption{SM Higgs inflation in the $(\ns,M_t)$ plane. Contours (68/95\%)
are the joint data constraint, $\ns$ posterior times the Top-mass
Gaussian, for CMB (blue) and CMB+DESI (orange).
The SM-Higgs posterior (indigo; CMB$\,\times\,M_t$, conditional on inflation) and
its full $\Nstar=50$--$60$ reach (faint band) sit below the inflation cliff
$M_t\simeq170.6$~GeV (dashed, with a $\pm1$~GeV theory band). The arrow marks the $\simeq3\sigma$ tension.}
\label{fig:higgsnsmt}
\end{figure}
The SM Higgs inflation with the
RG-improved one-loop potential has no tunable deviation such as $t$ or $c$ of the previous models. Its closest analogue is the top mass $M_t$, on which the
one-loop correction depends through the running couplings. We therefore treat
$M_t$ as a free parameter. 
At least at the 1-loop level, the $\ns$ value is actually decreased. The largest $\ns$ over the entire physical box and for every $M_t$ is 
$\Nstar=60,\,\ns\simeq0.969$, 
never reaching the
DESI-preferred value $0.9728$. 
Moreover, Inflation bounds the allowed top mass $M_t\lesssim170.8$~GeV. A higher value of $M_t$ and the RG flow will cause the quartic coupling $\lambda$ to become negative \cite{Buttazzo:2013uya}, precluding inflation without additional UV completion. 
Specifically, the measured value $M_t=172.57\pm0.29$~GeV lies beyond the inflationary window. 
This can be made quantitative by folding in the top-mass measurement
with a broadened standard deviation $\sigma_{\rm eff}\sim\!1$~GeV due to theory uncertainty~\cite{Degrassi:2012ry}, 
resulting in a Gaussian likelihood
$\mathcal{N}(M_t;172.57,\sigma_{\rm eff})$. Multiplied into the CMB $(\ns,\rr)$
likelihood, it gives a joint constraint in the $(\ns,M_t)$ plane, Fig.~\ref{fig:higgsnsmt}. Since this confronts a {\it
single} model with data, the relevant statistic is a tension rather than a Bayes
factor: the model reaches neither the $\ns$ nor the $M_t$ the data prefer, a joint
$(\ns,M_t)$ tension of $\simeq3\sigma$. 
\newline
\textbf{Conclusions:} Treating inflationary potentials at tree level is no longer adequate
with current and future precision. Incorporating quantum corrections
moves predictions of otherwise-successful models by an observable
amount, in a direction set by the underlying theory. The present preference is
modest and driven by the DESI-induced upward shift in $\ns$. A comparable improvement in $\ns$ determination from CMB-S4-class experiments or tighter BAO would sharpen it into either a detection, or a stringent bound on the
quantum corrections. 
A full-likelihood reanalysis and a joint treatment of the
scalar amplitude would further test the robustness of the hint.
\textbf{Acknowledgments:} We acknowledge the Ariel HPC Center at Ariel University for providing computing resources that have contributed to the research results reported within this paper.

\bibliographystyle{apsrev4-2}
\bibliography{ref}

\appendix

\section{Data and analysis details}
\label{app:method}

\subsection{Likelihood from marginalized posteriors}

Because the model modifies {\it only} the primordial sector, its entire
observational leverage is carried by the predicted $(n_s,r)$. We therefore
build the likelihood directly from the publicly released, fully marginalized
$(n_s,r)$ posterior of each dataset, represented as a
weighted Gaussian kernel density estimate (KDE),
\begin{equation}
\ln\mathcal{L}(t,N_{\star})=\ln\mathrm{KDE}_{n_s,r}\!\big(n_s(t,N_{\star}),\,r(t,N_{\star})\big).
\label{eq:likeli}
\end{equation}
Since the public chain already marginalizes the full $\Lambda$CDM and foreground
parameter space, and the model changes no (or negligible) parameter outside the primordial
$(n_s,r)$ sector, the marginalized $(n_s,r)$ posterior is a sufficient
statistic for it. Using Eq.~\eqref{eq:likeli} is statistically equivalent to a full
re-analysis at a small fraction of the cost. 



\subsection{Priors, sampling, and convergence}

The $R^{2}\ln R$ posterior is sampled over $(t,\Nstar)$ with the adaptive
Metropolis sampler of \textsc{Cobaya}~\cite{Cobaya}, under flat priors
$t\in[0,0.035]$ and $\Nstar\in[50,60]$. Convergence is monitored with the
Gelman--Rubin statistic and the run is stopped at $R-1<10^{-3}$; the chains reach
$R-1=8.1\times10^{-4}$ (CMB) and $7.3\times10^{-4}$ (CMB+DESI) at a mean Metropolis
acceptance rate $\simeq0.28$. After discarding the first $30\%$ of each chain as
burn-in, they retain $\simeq1.6\times10^{4}$ and $\simeq2.4\times10^{4}$ weighted
samples, corresponding to effective sample sizes $\simeq1.4\times10^{3}$ and
$\simeq2.4\times10^{3}$. Marginalized posteriors are processed with
\textsc{GetDist}~\cite{GetDist}, whose boundary correction is applied at the $t=0$
edge (the posterior rails there for CMB alone). The non-minimal scalar is sampled over flat
priors $c\in[0,0.035]$ and $\Nstar\in[50,60]$, the same numerical ranges as the
$R^{2}\ln R$ model, so the two deformation parameters are given equal-width priors
for a prior-fair comparison.

\subsection{Model-comparison statistics}
\label{sec:stats}

Pure Starobinsky ($\Mzero$: zero deviation) is nested at the prior boundary of
each one-parameter extension ($\Mone$: $t>0$ for $R^{2}\ln R$, or $c>0$ for the
non-minimal scalar). We quantify the preference with four complementary
statistics, applied identically to both extensions and summarized in
Table~\ref{tab:modelcomp}:

\paragraph{Profile $\Delta\chi^{2}$.} The exact difference
$\Delta\chi^{2}=\chi^{2}(t{=}0,\hat N)-\chi^{2}(\hat t,\hat N)$ between the
best-fit one-loop point and the best Starobinsky point (profiled over $\Nstar$),
where $\chi^{2}\equiv-2\ln\mathcal{L}$. Because $t=0$ is a boundary, the
statistic follows a $\tfrac12\chi^{2}_{0}+\tfrac12\chi^{2}_{1}$ mixture and the
quoted significance is one-sided.

\paragraph{Savage--Dickey density ratio.} For the nested point $t=0$ with
separable, uniform priors, the Bayes factor is
$B_{01}=p(t{=}0\,|\,d)/\pi(t{=}0)$, with the posterior density estimated by a
boundary-corrected KDE.

\paragraph{Bayesian evidence.} We compute the evidences
$Z_{1}=\int\mathcal{L}\,\pi(t)\pi(\Nstar)\,dt\,d\Nstar$ and
$Z_{0}=\int\mathcal{L}(0,\Nstar)\,\pi(\Nstar)\,d\Nstar$ by dense two-dimensional
quadrature (the gold standard for this low-dimensional problem) and
independently obtain $Z_{1}$ with a self-contained Skilling nested
sampler~\cite{Skilling2006} (400 live points) as a cross-check. We report
$\lnB=\ln Z_{0}-\ln Z_{1}$.

\paragraph{Deviance information criterion.} As a {\it prior-independent} measure
we use $\mathrm{DIC}=\bar D+p_{D}$ with $\bar D=\langle-2\ln\mathcal{L}\rangle$,
$p_{D}=\bar D-D(\bar\theta)$~\cite{Spiegelhalter2002,GelmanBDA}. 
The
Starobinsky and non-minimal DICs are obtained exactly by quadrature over their
posteriors---one-dimensional over $\Nstar$ at $t=0$ for Starobinsky, and
two-dimensional over $(c,\Nstar)$ for the non-minimal scalar---requiring no
separate chain, while the $R^{2}\ln R$ DIC uses its MCMC chain. We quote
$\Delta\mathrm{DIC}=\mathrm{DIC}(\Mone)-\mathrm{DIC}(\Mzero)$ for each extension.

For both Bayesian measures, $\lnB>0$ and $\Delta\mathrm{DIC}>0$ favor
Starobinsky; negative values favor the one-loop extension. We interpret
$\lnB$ on the Jeffreys/Kass--Raftery scale~\cite{KassRaftery1995} and
$|\Delta\mathrm{DIC}|$ on the usual rule of thumb ($<2$ negligible, $2$--$5$
positive, $5$--$10$ strong). Hence, we get positive evidence for both $R^{2}\ln R$, and non-minimal model.

\section{Marginalized posteriors and supplementary figures}
\label{app:post}

\begin{figure}[!tb]
\centering
\includegraphics[width=\columnwidth]{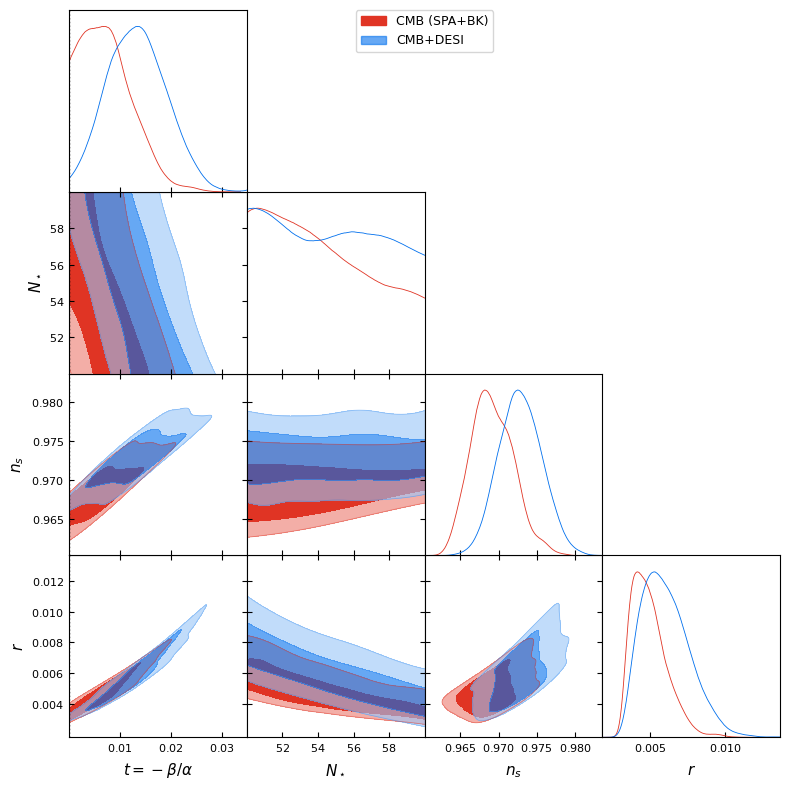}
\caption{Marginalized one- and two-dimensional posteriors of
$(t,\Nstar,\ns,\rr)$ for CMB (blue) and CMB+DESI (red). The dotted line marks the
Starobinsky boundary $t=0$. $\Nstar$ is prior-dominated.}
\label{fig:triangle}
\end{figure}

\begin{figure}[!tb]
\centering
\includegraphics[width=0.85\columnwidth]{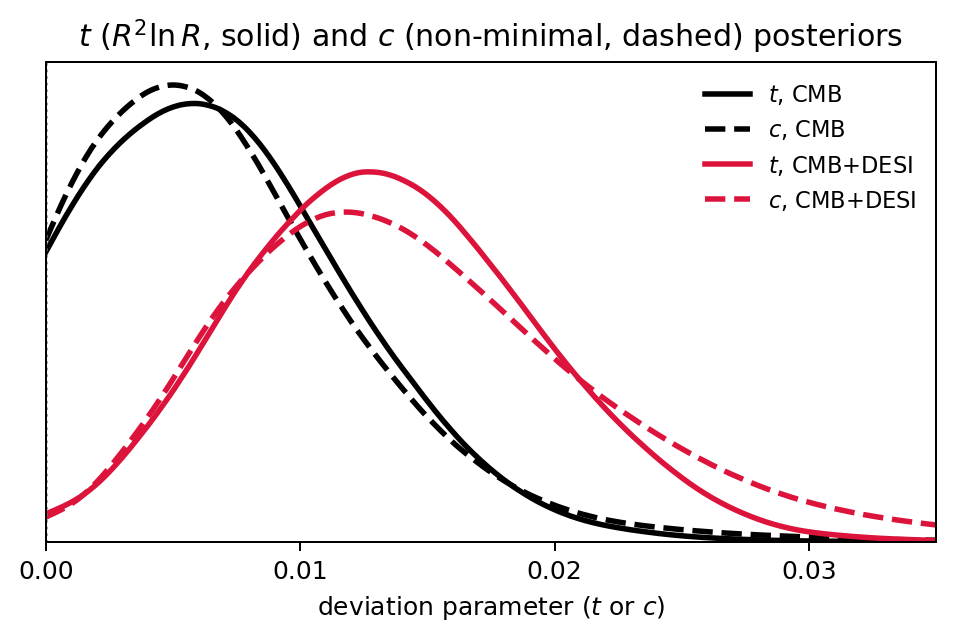}
\caption{Marginalized posteriors of the correction strengths $t=-\beta/\alpha$
($R^{2}\ln R$, solid) and $c=9\lambda/8\sqrt{6}\,\pi^{2}$ (non-minimal scalar,
dashed), for CMB (black) and CMB+DESI (red); zero is pure
Starobinsky.}
\label{fig:t1d}
\end{figure}

\begin{figure}[!tb]
\centering
\includegraphics[width=\columnwidth]{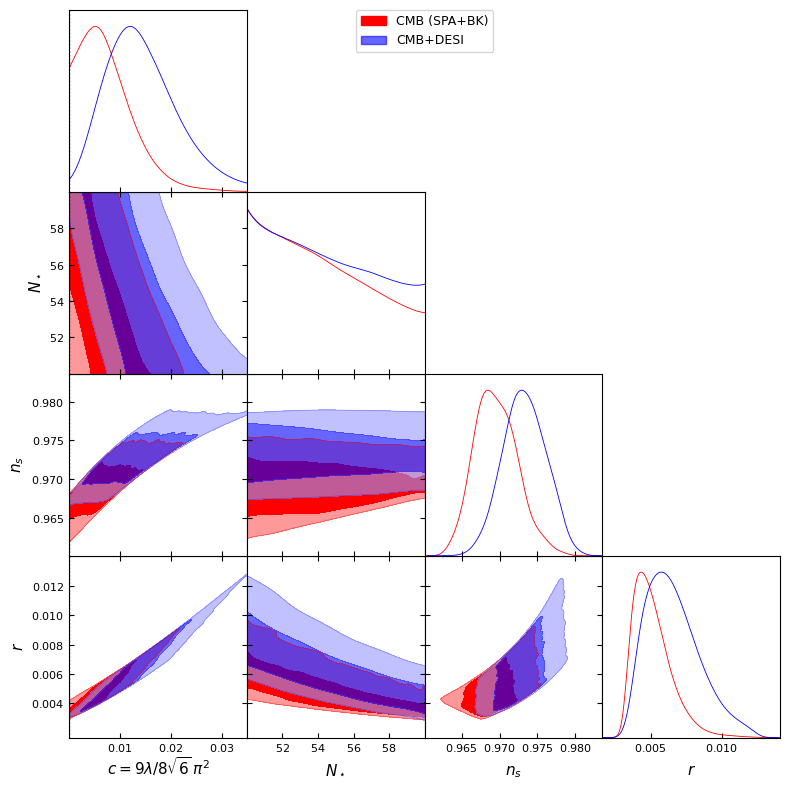}
\caption{Marginalized posteriors of the non-minimal scalar,
$(c,\Nstar,\ns,\rr)$, for CMB
(red) and CMB+DESI (blue). CMB prefers a
small tilt, predictions sitting near the Starobinsky point, while CMB+DESI
prefers a larger tilt $c\approx0.014$ ($\lambda\approx0.3$). $\Nstar$ is
prior-dominated, reflecting the $c$--$\Nstar$ degeneracy.}
\label{fig:nmtriangle}
\end{figure}

This appendix collects the full marginalized posteriors and the auxiliary
$(\ns,\rr)$ figures underlying the main-text results: the $(\ns,\rr)$ constraint
plot with 
the $R^{2}\ln R$ and
non-minimal posterior triangles (Figs.~\ref{fig:triangle} and
\ref{fig:nmtriangle}), and the one-dimensional correction-strength posterior
(Fig.~\ref{fig:t1d}). It is clear that CMB+DESI prefer the deviation from tree-level, and the inclusion of a quantum correction at the level of nearly $2\sigma$, see Fig.~\ref{fig:t1d}.


\end{document}